\documentclass[superscriptaddress, reprint, amsmath, amssymb, aps, pra, floatfix]{revtex4-2}

\usepackage{bm}
\usepackage{algpseudocode}
\usepackage{algorithm}

\usepackage{mathtools}
\usepackage{amsfonts}
\usepackage{amssymb}
\usepackage{dsfont}
\usepackage{amsthm}
\usepackage[english]{babel}
\usepackage{graphicx}
\usepackage{amsmath}
\usepackage{lipsum}
\usepackage{wrapfig}
\usepackage{float}
\usepackage{enumitem} 
\usepackage{color}
\usepackage{hyperref}
\usepackage[normalem]{ulem}
\usepackage{url}

\useunder{\uline}{\ul}{}

\hypersetup{colorlinks = true}

\def\bra#1{\ensuremath{\mathinner{\langle{#1}|}}}
\def\ket#1{\ensuremath{\mathinner{|{#1}\rangle}}}

\newcommand{\needcite}[1]{\textcolor{red}{[Ref needed]}}
\usepackage{qcircuit}

\begin{document}

\title{Randomized adaptive quantum state preparation}

\author{Alicia B. Magann}
\affiliation{Quantum Algorithms and Applications Collaboratory, Sandia National Laboratories, Albuquerque, New Mexico 87185, USA}

\author{Sophia E. Economou}
\affiliation{Department of Physics, Virginia Tech, Blacksburg, Virginia 24061, USA}

\author{Christian Arenz}
\affiliation{School of Electrical, Computer, and Energy Engineering, Arizona State University, Tempe, Arizona 85287, USA}

\date{\today}

\begin{abstract}
We develop an adaptive method for quantum state preparation that utilizes randomness as an essential component and that does not require classical optimization. Instead, a cost function is minimized to prepare a desired quantum state through an adaptively constructed quantum circuit, where each adaptive step is informed by feedback from gradient measurements in which the associated tangent space directions are randomized. We provide theoretical arguments and numerical evidence that convergence to the target state can be achieved for almost all initial states. We investigate different randomization procedures and develop lower bounds on the expected cost function change, which allows for drawing connections to barren plateaus and for assessing the applicability of the algorithm to large-scale problems.

\end{abstract}

\maketitle

\section{Introduction}
Methods for preparing quantum states are an integral component of any quantum technology. For example, the preparation of states that encode ground, excited, and thermal states of many-body systems is a key element in quantum simulation \cite{RevModPhys.92.015003,cao2019quantum}. Ground state preparation can also be leveraged to solve combinatorial optimization problems, with a variety of applications including in routing and scheduling \cite{Papadimitriou_Steiglitz_1998}. Although the task of ground state preparation is known to be hard, including for quantum computers, there is nonetheless significant interest in algorithms for quantum state preparation \cite{Kitaev2002May, kempe2006complexity}. In particular, the growing availability of noisy, intermediate-scale quantum \cite{preskill_quantum_2018} devices has inspired immense interest in variational methods for preparing desired quantum states \cite{peruzzo2014variational, cerezo2021variational,  https://doi.org/10.48550/arxiv.1411.4028, PRXQuantum.2.010101}. These variational quantum algorithms (VQAs) are heuristics that function by classically optimizing over a set of parameters that enter into a quantum circuit whose structure is typically fixed. The parameterized quantum circuit is executed on a quantum device and serves as an ansatz to minimize a cost function, $J$, whose global minimum is achieved for the desired target state. 

Even in the absence of noise, a variety of challenges are present in VQAs. On the quantum device, for example, one must select a quantum circuit ansatz and associated initial state out of a formidably large design space. Meanwhile, the difficulty of the cost function minimization means that the challenges on the classical side can be even more significant \cite{bittel2021training}, often involving the navigation of optimization landscapes that contain barren plateaus and suboptimal local minima.

\begin{figure}[!t] 
\centering
\includegraphics[width=0.8\columnwidth]{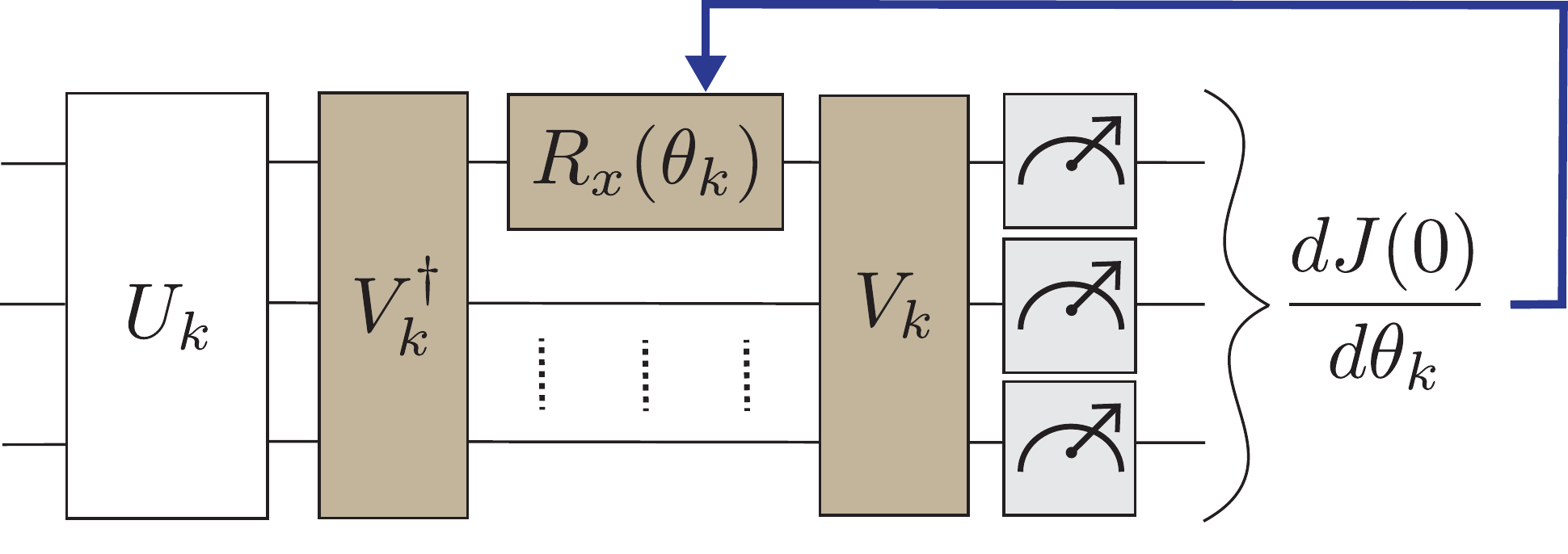}
\caption{Schematic representation of randomized adaptive quantum state preparation. Each adaptive step $k$ involves first estimating the gradient $\frac{dJ(0)}{d\theta_{k}}$ at $\theta_{k}=0$ of a cost function $J$ with respect to a parameter $\theta_{k}$, e.g., a rotation angle, whose corresponding direction is randomized through conjugation with a random unitary transformation $V_{k}$ (brown). The parameter $\theta_{k}$ is then updated in the negative direction of the gradient (blue arrow), which moves the system closer to the target state, and the 
quantum circuit $U_{k}$ is extended accordingly.}
\label{Fig:Intro}
\end{figure}
 
To overcome these challenges, approaches have been proposed that utilize feedback from qubit measurements to \emph{adaptively} construct a quantum circuit to minimize $J$. The first such algorithm was the Adaptive Derivative-Assembled Problem-Tailored Variational Quantum Eigensolver (ADAPT-VQE) \cite{grimsley2019adaptive, Tang2021PRXQ}, which was also adapted to combinatorial optimization problems \cite{PhysRevResearch.4.033029}. Instead of relying on a predefined ansatz, ADAPT-VQE grows it in a layer-wise manner tailored to the problem, in tandem with classical optimization over the quantum circuit parameters.

Other methods have considered defining the structure of the ansatz \emph{a priori} and then performing layer-wise optimization to adaptively set the circuit parameter values \cite{carolan2020variational, skolik2021layerwise,PhysRevA.103.032607,campos2021training}. Adaptive methods that do not require any classical optimization have also been developed, including the feedback-based algorithm for quantum optimization \cite{falqon1,falqon2,larsen2023feedbackbased} and methods based on Riemannian gradient flows \cite{https://doi.org/10.48550/arxiv.2202.06976,doi:10.1142/S0129055X10004053}. In methods where each adaptive step is efficiently implementable, e.g., \cite{grimsley2019adaptive, Tang2021PRXQ,PhysRevResearch.4.033029,carolan2020variational, skolik2021layerwise,PhysRevA.103.032607,campos2021training,falqon1,falqon2}, and the approximate method in \cite{https://doi.org/10.48550/arxiv.2202.06976}, the circuit growth can get stuck when the gradient vanishes. This means that these methods can face similar issues as conventional VQAs that are prone to converge to suboptimal solutions.

Here, we propose \emph{randomized} adaptive quantum state preparation as a generic, adaptive quantum algorithm that minimizes up-front design choices, does not require classical optimization, and allows for preparing target quantum states from arbitrary (random) initial states. As a consequence of the latter point, the algorithm, or its key subroutine depicted in Fig. \ref{Fig:Intro}, can be readily combined with other state preparation methods to improve convergence and state preparation fidelities. We provide theoretical arguments and numerical evidence that substantiate our claims about convergence, explore different methods for achieving randomization in practice, and develop lower bounds on the expected change in $J$ at each adaptive step. These bounds give guarantees for how much the cost function value changes when randomization is used, thereby allowing us to relate the efficiency of this randomized approach to the existence of barren plateaus \cite{mcclean2018barren}.  We go on to discuss how this approach can be applied to cooling in open quantum systems and mixed state preparation in general.

\section{Randomized adaptive quantum algorithms}

We consider minimizing cost functions of the form 
\begin{align}
    J_k = \bra{\psi_{k}}H_{p}\ket{\psi_{k}},
    \label{Eq:J}
\end{align}
by creating states $\ket{\psi_{k}}=U_{k}\ket{\psi_{0}}$ via an adaptively constructed quantum circuit $U_{k}$, $k=0,1,\cdots$. The goal is to apply this circuit to a fixed initial state $\ket{\psi_{0}}$ to achieve $J_{k+1}\leq J_{k}$ in each adaptive step $k$. Here, $H_{p}$ is a Hermitian operator whose ground state $\ket{E_{\text{min}}}$, with corresponding eigenvalue $E_{\text{min}}$, is taken to be the target state. We note that in general, knowledge of the initial and target states $\ket{\psi_{0}}$ and $\ket{E_{\text{min}}}$ is not required. However, for the preparation of an arbitrary, known target state $\ket{\psi_{T}}$, $H_{p}$ can be defined as $H_p = \mathds{1} - \ket{\psi_{T}}\bra{\psi_T}$.  

We consider a quantum circuit that is adaptively created according to 
\begin{align}
    U_{k+1}=e^{-i\theta_{k}H_{k}}U_{k},~~~~U_{0}=\mathds{1},
    \label{Eq:AdaptiveUpdate}
\end{align}
where in each step $k$ we move into the negative direction of the gradient of $J$ with respect to $\theta_k$ by setting $\theta_{k}\leftarrow -\gamma\frac{dJ(0)}{d\theta_{k}}$, where $\frac{dJ(0)}{d\theta_{k}}=i\bra{\psi_{k}}[H_{k},H_{p}]\ket{\psi_{k}}$ is the gradient of $J(\theta_{k})=\bra{\psi_{k}}e^{i\theta_{k}H_{k}}H_{p}e^{-i\theta_{k}H_{k}}\ket{\psi_{k}}$ evaluated at $\theta_{k}=0$. Alternatively, in situations where $J_k$ can be estimated via repeated measurements of $H_p$ (e.g., for $H_p = \mathds{1} - \ket{\psi_{T}}\bra{\psi_T}$ with known $|\psi_T\rangle$), the derivative $\frac{dJ(0)}{d\theta_{k}}$ can be estimated via a finite difference approximation, by estimating $J_k$ for different perturbations of $\theta_k$. For sufficiently small learning rates $\gamma$, this ensures that $J_{k+1}\leq J_{k}$. References \cite{grimsley2019adaptive,PhysRevResearch.4.033029,falqon1,falqon2,https://doi.org/10.48550/arxiv.2202.06976} consider adaptive procedures similar to Eq. \eqref{Eq:AdaptiveUpdate}, and in cases where the circuit growth gets stuck in suboptimal solutions, i.e., when the gradient vanishes, the utility of incorporating randomness into the circuit structure to escape these suboptima has been observed numerically in \cite{https://doi.org/10.48550/arxiv.2202.06976}.

In this work, we utilize randomness to overcome challenges associated with convergence through the introduction of an \emph{intrinsically randomized} framework for quantum state preparation in which the $H_{k}$'s are selected at random. In the following, we provide theoretical arguments that this randomization enables convergence from almost all initial states to arbitrary target states.

We first note that the cost function gradient can be expressed as
\begin{align}
    \frac{dJ(0)}{d\theta_{k}}=\langle \text{grad}J[U_{k}],iH_{k}\rangle,
    \label{Eq:Gradient}
\end{align} 
where $\langle \cdot,\cdot\rangle$ denotes the Hilbert-Schmidt inner product and $\text{grad} J[U_{k}]=[\ket{\psi_{k}}\bra{\psi_{k}},H_{p}]$ is the (Riemannian) gradient (up to multiplication with $U_{k}$ from the right) of $J[U_{k}] = \langle\psi_0|U_k^\dagger H_p U_k|\psi_0\rangle$ with respect to the unitary transformation $U_{k}$ \cite{doi:10.1142/S0129055X10004053, helmke2012optimization, absil2009optimization}. Both $iH_{k}$ and $\text{grad} J[U_{k}]$ belong to the special unitary algebra $\mathfrak{su}(2^{n})$ consisting of all traceless and anti-Hermitian $2^{n}\times 2^{n}$ matrices where $n$ is the number of qubits. From this geometric perspective we can now deduce two different cases (i) and (ii) for when $\frac{dJ(0)}{d\theta_{k}}$ vanishes \footnote{We neglect the trivial third case that $iH_k=0$.}. In case (i), the gradient vanishes when $iH_{k}$ is orthogonal to $\text{grad} J[U_{k}]\neq 0$, while in case (ii) $\text{grad} J[U_{k}]=0$, which happens when $H_{p}$ commutes with $\ket{\psi_{k}}\bra{\psi_{k}}$. 

We first discuss case (i). Typically, no assumptions can be made on whether $\text{grad} J[U_{k}]$ moves into a lower dimensional subspace of $\mathfrak{su}(2^{n})$ when growing the circuit. As such, the situation that $iH_{k}$ has no overlap with $\text{grad} J[U_{k}]$ can occur, e.g., when $iH_k$ is an element of a subspace over which $\text{grad} J[U_{k}]$ has no support. To overcome this issue, we propose to create each $H_{k}$ at random. This can be achieved by conjugating a traceless Hermitian operator $H$ by a Haar random unitary transformation $V_{k}$ in each adaptive step, i.e., such that $H_{k}=V_{k}^{\dagger}HV_{k}$.  Since $H_{k}$ is created uniformly randomly according to the Haar measure, the probability that $iH_{k}$ is orthogonal to $\text{grad} J[U_{k}]$ is zero. That is, for almost all $iH_{k}$, but a set of measure zero, case (i) does not occur. While Haar random unitaries are not efficiently implementable, below we discuss the efficient implementation via approximate unitary 2-designs \cite{PhysRevA.80.012304}.

We now focus on case (ii). For cost functions of the form \eqref{Eq:J}, the set of critical points where $\text{grad} J[U_{k}]$ vanishes consists of global optima and saddle points only \cite{BROCKETT1989761,Wu_2008}. Under mild assumptions on the nature of the saddle points (strict saddles), relevant works from the classical machine learning and optimization literature have found that saddle points are avoided for almost all initial conditions \cite{pmlr-v40-Ge15, pmlr-v70-jin17a, almostalwayssaddle}. We thus expect that randomized adaptive quantum state preparation will almost surely converge to the ground state of $H_{p}$ for almost all initial states $\ket{\psi_{0}}$. We remark that the convergence result cannot hold for \emph{all} initial states, as we immediately see that for eigenstates of $H_{p}$, $\text{grad}J[U_{0}]=0$. However, the situation that $\text{grad}J[U_{0}]=0$ can be avoided with probability one when the initial state is randomized too. 

Each step $k$ of randomized adaptive quantum state preparation can now be summarized as follows. First, the unitary transformation $V^{\dagger}_{k}e^{-i\theta_{k}H}V_{k}$, whose generator $H$ is randomized through conjugation with a random unitary $V_{k}$, is applied to the state $\ket{\psi_{k}}$. Second, the gradient, $\frac{dJ(0)}{d\theta_{k}}$, is estimated, e.g., using the parameter shift rule \cite{PhysRevLett.118.150503,PhysRevA.98.032309,PhysRevA.99.032331,Wierichs2022generalparameter,PhysRevA.104.052417}. Third, the parameter $\theta_{k}$ is updated in the negative direction of the gradient, i.e., $\theta_{k}\leftarrow -\gamma\frac{dJ(0)}{d\theta_{k}}$.

\section{Efficiency and relation to barren plateaus}

Randomized adaptive quantum state preparation is not expected to be efficient in general, as ground state preparation is QMA complete \cite{kitaev2002classical, doi:10.1137/S0097539704445226}. Here, we show that for a given problem, the efficiency can be related to the scaling of the gradient \eqref{Eq:Gradient} with the system size, and therefore, to the existence of barren plateaus \cite{mcclean2018barren}, i.e., exponentially flat regions in the optimization landscape where the variance of the gradient vanishes exponentially in the number of qubits $n$.

In Appendix \ref{sec:appA}, we show that if we select $\gamma=1/(4\Vert H_{p}\Vert_{2})$ and assume $\Vert H_{k} \Vert_{2} =1$, where $\Vert \cdot \Vert_{2} $ denotes the spectral norm, then the cost function change $\Delta J_{k}=J_{k}-J_{k+1}$ is lower bounded by 
\begin{align}
    \Delta J_{k}\geq \frac{1}{8\Vert H_{p}\Vert_{2}}\left(\frac{dJ(0)}{d\theta_{k}}\right)^{2}.
    \label{Eq:ChangeinJ}
\end{align}
If we assume that it takes $M$ steps to create the ground state up to an error $\epsilon$, i.e., $J_{M}=E_{\text{min}}+\epsilon$, we find from \eqref{Eq:ChangeinJ} that $M$ is upper bounded by, 
\begin{align}
M\leq \frac{C_{\epsilon}}{\min\limits_{k\in [0,M]}\langle \text{grad}J[U_{k}],iH_{k}\rangle^{2}}, 
\label{Eq:MUpperBound}
\end{align}
where the constant in the numerator is given by $C_{\epsilon}=8\Vert H_{p}\Vert_{2}\big(J_0-(E_{\text{min}}+\epsilon)\big)$. 
Thus, if $\langle\text{grad}J_{k}[U_{k}],iH_{k}\rangle^{2}$ does not vanish faster than $1/\text{poly}(n)$,  the ground state of $H_{p}$ can be prepared up to precision $\epsilon$ (in the corresponding eigenvalue) in polynomially many steps. We remark that the minimum in the denominator of Eq. \eqref{Eq:MUpperBound} explicitly depends on $M$, and thereby on the random path taken. It is interesting to note that similar expressions are obtained in adiabatic state preparation \cite{https://doi.org/10.48550/arxiv.quant-ph/0001106}, where the scaling of the adiabatic state preparation time $T$ is determined by the smallest value of the spectral gap $\Delta(t)$, taken over all times, i.e., $\min_{t\in[0,T]}\Delta(t)$ \cite{RevModPhys.90.015002}.

\begin{figure}[!t] 
\centering
\includegraphics[width=0.85\columnwidth]{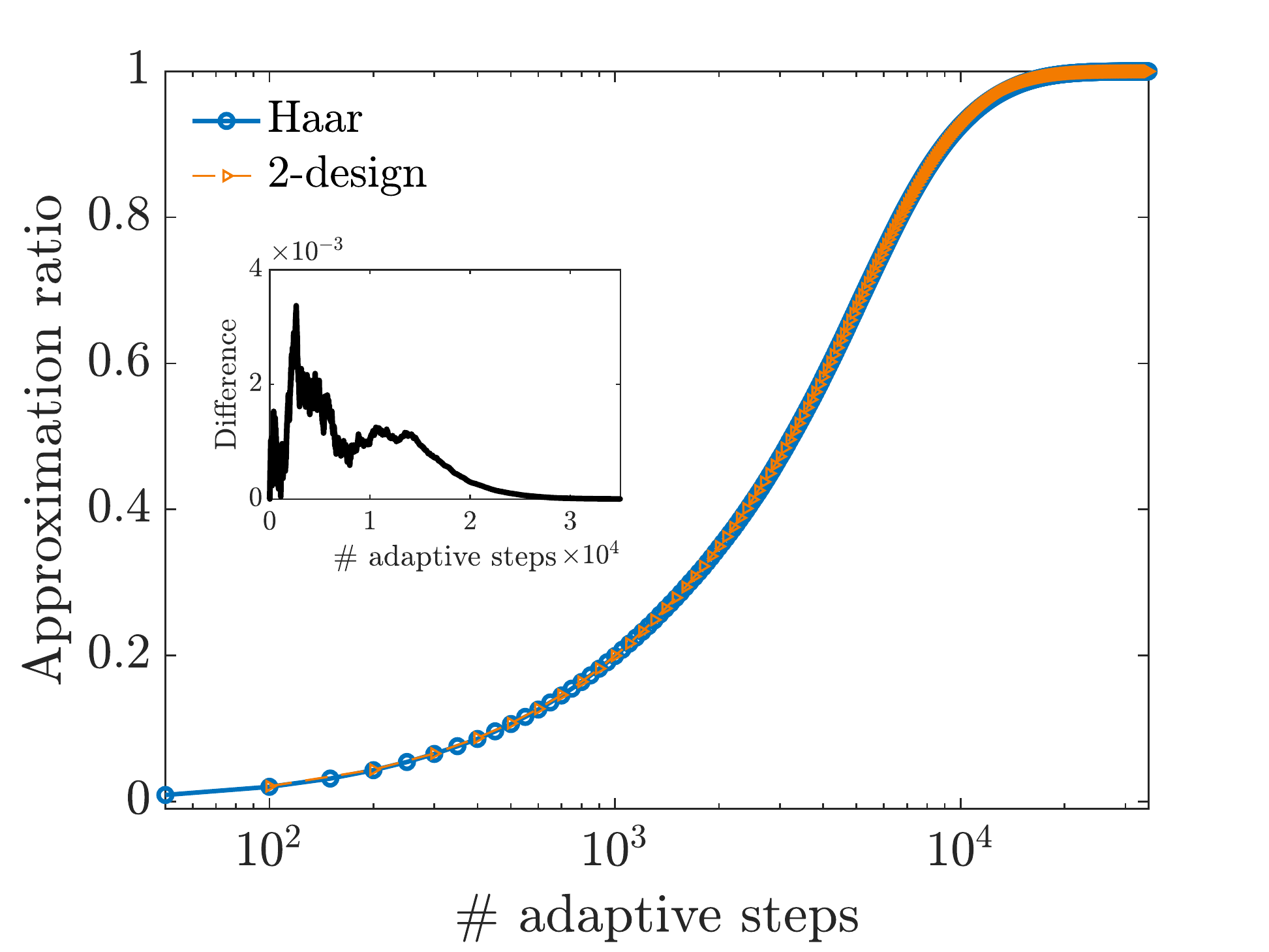}
\caption{Performance of randomized adaptive quantum state preparation when $H_p$ is an Ising Hamiltonian with $n=8$ spins. We map each spin to a vertex on a 3-regular graph with 8 vertices, and couple spins whose corresponding vertices are connected by an edge. Each data point corresponds to the average of the approximation ratio, $\alpha$, taken over $100$ different algorithm realizations and initial states. This is plotted as a function of the number of adaptive steps, $M$, on a logarithmic scale. In each step, the randomization of $H_{k}$ is implemented by conjugating the Pauli operator $X_1I_2\cdots I_n$ with a Haar random unitary transformation (blue circles) and with a unitary transformation sampled from an approximate unitary 2-design (orange triangles), as depicted in Fig. \ref{Fig:Intro}, created using the sequence in \cite{doi:10.1063/1.4983266} with $\ell=1$. The two curves are nearly superimposed. Inset shows the difference between the two curves, computed as $|\alpha_{\text{Haar}}-\alpha_{\text{2-design}}|$, which never exceeds $4\times 10^{-3}$.}
\label{Fig:performance3regulr}
\end{figure}

We proceed by investigating the efficiency of different randomization strategies. At each step $k$, the expected cost function change $\mathbb E_{H_{k}}\Delta J_{k}$ is lower bounded by the variance $\mathbb E_{H_{k}}\langle\text{grad}J[U_{k}],iH_{k}\rangle^{2}$ of the gradient, up to the prefactor in (\ref{Eq:ChangeinJ}), assuming that the expectation vanishes. This suggests that sampling $V_{k}$ from unitary 2-designs suffices to obtain convergence to the ground state. This observation is further substantiated by the numerical simulations in Fig. \ref{Fig:performance3regulr}, which considers the task of preparing the ground state of an Ising Hamiltonian. We specifically consider a model in which we map each spin to a vertex on a 3-regular graph, and couplings are present between spins whose corresponding vertices are connected by an edge. This is equivalent to solving the combinatorial optimization problem MaxCut on an unweighted, 3-regular graph \cite{barahona1988application}. We consider the approximation ratio $\alpha=J_k/E_\text{min}$ as our figure of merit. Fig. \ref{Fig:performance3regulr} compares results of randomized adaptive quantum state preparation when $V_k$ is sampled at random from the Haar measure, with results when $V_k$ is sampled from a unitary 2-design. The results that are obtained are nearly identical, with the difference shown in the inset.

The appearance of the variance in the bound \eqref{Eq:ChangeinJ} means that we additionally expect the efficiency and practical utility of this method to be closely related to the existence of barren plateaus. While barren plateaus pose a major challenge to the scalability of VQAs \cite{mcclean2018barren,PRXQuantum.2.040316,wang2021noise}, examples have been found where the variance of the gradient does not vanish faster than $1/\text{poly}(n)$ \cite{BarrenPoly}. Leveraging these instances for efficient realizations of randomized adaptive quantum state preparation will be the subject of future studies.

We now consider lower bounds for $\mathbb E_{H_{k}}\Delta J_{k}$ to derive guarantees for how much $J_k$ can be improved when randomization is used. When $H_{k}$ is created through conjugation by a unitary $V_{k}$ sampled from a unitary 2-design, we show in Appendix \ref{sec:appB} that $\mathbb E_{V_{k}}\Delta J_{k}$ is lower bounded by
\begin{align}
\label{eq:lowerboundHaar}
\mathbb E_{V_{k}}\Delta J_{k} \geq \frac{\text{Tr}\{H^{2}\}}{4\Vert H_{p}\Vert_{2}}\frac{\text{Var}_{\psi_{k}}(H_{p})}{2^{2n}-1},    
\end{align}
where $\text{Var}_{\psi_{k}}(H_{p})=\bra{\psi_{k}}H_{p}^{2}\ket{\psi_{k}}-\bra{\psi_{k}}H_{p}\ket{\psi_{k}}^{2}$ is the variance of $H_p$ with respect to the state $\ket{\psi_{k}}$. Since $\text{Var}_{\psi_{k}}(H_{p})$ is bounded from above by a constant that is independent of the system dimension, we see that the bound in (\ref{eq:lowerboundHaar}) vanishes exponentially in $n$.

Another way of creating random $H_{k}$'s is by sampling uniformly from an operator pool $\mathcal A$ whose size we denote by $|\mathcal A|$. While in this case, situation (i) can occur when an operator is selected that is orthogonal to $\text{grad}J_{k}[U_{k}]$, on average, we have that
\begin{align}
\mathbb E_{H_{k}\in\mathcal A} \Delta J_{k}\geq \frac{1}{8\Vert H_{p}\Vert_{2} |\mathcal A|}\sum_{H_{k}\in\mathcal A}\langle \text{grad}J[U_{k}],iH_{k}\rangle^{2},
\label{eq:pool}
\end{align}
which suggests that situation (i) can be avoided on average for sufficiently large operator pools, e.g., when $\text{span}\{\mathcal A
\}=\mathfrak{su}(2^n)$. Note that in this case, convergence to the ground state can also be obtained by simply continuing to sample from $\mathcal A$ when (i) occurs. However, in this situation we expect an exponential runtime, as $M$ scales as $1/|\mathcal A|$.  For $|\mathcal A| = \text{poly}(n)$, on the other hand, convergence to the ground state is no longer guaranteed, as the adaptive procedure can get stuck in suboptima where the gradient vanishes, due to (i). Furthermore, while the situation that $|\mathcal A| = \text{poly}(n)$ implies polynomial scaling of the denominator of Eq. (\ref{eq:pool}), it does not necessarily imply that randomized adaptive quantum state preparation would be efficient in this setting, as it does not imply polynomial scaling of the numerator.

\begin{figure}[!t] 
\centering
\includegraphics[width=0.85\columnwidth]{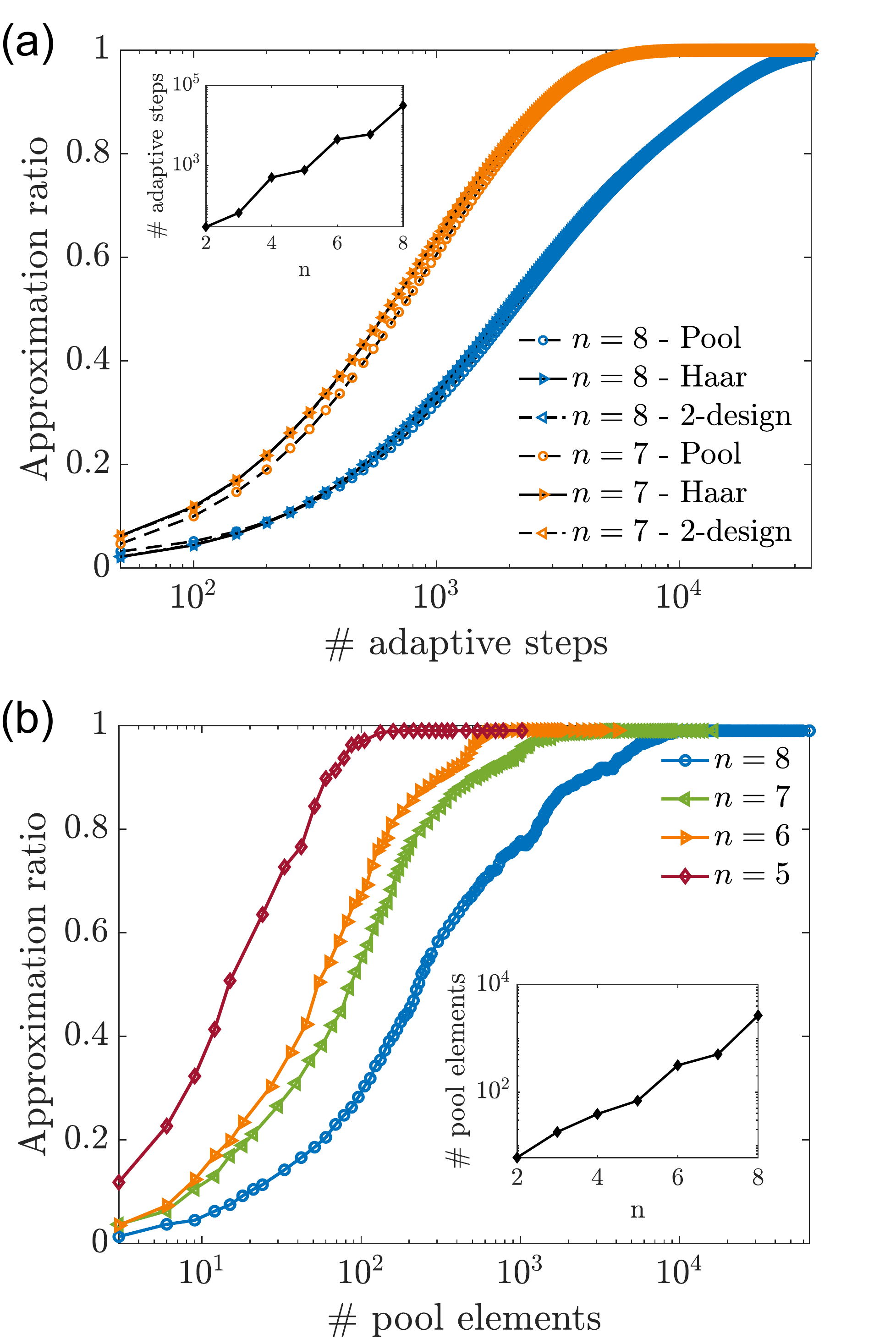}
\caption{Performance of randomized adaptive quantum state preparation when $H_p$ is an Ising Hamiltonian with all-to-all couplings between $n$ spins. Each data point corresponds to the average taken over $100$ different algorithm realizations and initial states. In (a), the approximation ratio, $\alpha$, is plotted as a function of the number of adaptive steps, $M$, shown on a logarithmic scale. In each step, the randomization of $H_{k}$ is implemented by conjugating the Pauli operator $X_1I_2\cdots I_n$ with a Haar random unitary transformation (right triangles) and with a unitary transformation sampled from an approximate unitary 2-design (left triangles), as depicted in Fig. \ref{Fig:Intro}, created using the sequence in \cite{doi:10.1063/1.4983266} with $\ell=1$, and also by sampling $H_{k}$ uniformly from an operator pool $\mathcal A$ containing all $2^{2n}-1$ Pauli operators (circles). In (b), $\alpha$ is plotted as a function of the number of pool elements, $|\mathcal A|$, shown on a logarithmic scale, for fixed $M = 35,000$. Insets show semilog plots of the scaling with respect to $n$ for both cases to achieve $\alpha>0.99$ in (a) and $\alpha>0.9$ in (b).}
\label{Fig:performance}
\end{figure}

In Fig. \ref{Fig:performance}, we numerically investigate this tradeoff and study the convergence of randomized adaptive quantum state preparation for the problem of preparing the ground state of an Ising Hamiltonian with equal couplings present between all $n$ spins, which is equivalent to solving the combinatorial optimization problem MaxCut on an unweighted, complete graph \cite{barahona1988application}. In Fig. \ref{Fig:performance}(a), we plot the approximation ratio, $\alpha$, as a function of the number of adaptive steps $M$ for the different randomization strategies described above. In Fig. \ref{Fig:performance}(b), we consider fixed $M=35000$ and investigate $\alpha$ as a function of the pool size $|\mathcal A|$. The pool size is increased by adding successively heigher weight Pauli operators to $\mathcal{A}$, until the full pool used in Fig. \ref{Fig:performance}(a) containing all $2^{2n}-1$ terms is formed. We observe that the curves in Fig. \ref{Fig:performance}(a) are almost identical, suggesting that the three different randomization strategies converge in the same manner to the ground state for which $\alpha=1$. Furthermore, the curves in Fig. \ref{Fig:performance}(b) suggest that a full Pauli operator pool is not needed to obtain convergence to the ground state. However, the inset semilog plots in \ref{Fig:performance}(a) and \ref{Fig:performance}(b) do suggest an exponential scaling of the number of adaptive steps, $M$, and the number of operators in the pool, $|\mathcal{A}|$, with respect to $n$.

\section{Mixed states} 

Thus far, we have discussed the preparation of pure (ground) states from initial pure states. Here, we explore generalizations to preparing a target pure state from an initially mixed state, and vice versa. Since it is not possible to create pure quantum states from mixed quantum states in a closed quantum system through unitary transformations, we consider an extended, ``dilated'' space by coupling a set of system qubits, $S$, to a set of auxiliary qubits, $A$. We assume that the combined system is initially in a separable state $\rho_0^{SA} = \rho_0^S\otimes \rho_0^A$, where we denote the initial states of $S$ and $A$ by $\rho_{0}^{S}$  and $\rho_{0}^{A}$, respectively. We then consider growing a quantum circuit over the full composite system according to \eqref{Eq:AdaptiveUpdate}, in order to adaptively create the state $\rho_{k}^{SA}=U_{k}\rho_{0}^{SA}U_{k}^{\dagger}$. The state $\rho_{k}^{S}$ at the $k$-th adaptive step is then obtained by tracing over the degrees of freedom of the auxiliary qubits in subsystem $A$, i.e., $\rho_{k}^{S}=\text{Tr}_{A}\{\rho_{k}^{SA}\}$. 

In this setting, we now consider the task of preparing the system qubits in a target pure state $\ket{\psi_{T}}$. The cost function \eqref{Eq:J} becomes $J_k=1 - \bra{\psi_{T}}\rho_{k}^{S}\ket{\psi_{T}}$, and an adaptive change is described by $J(\theta_{k})=1-\bra{\psi_{T}}\text{Tr}_{A}\{e^{i\theta_{k}H_{k}}\rho_{k}^{SA}e^{-i\theta_{k}H_{k}} \} \ket{\psi_{T}}$. This yields the gradient 
\begin{align}
\label{eq:opensysgradient}
     \frac{dJ(0)}{d\theta_{k}}=-\langle [\rho_{k}^{SA},\ket{\psi_{T}}\bra{\psi_{T}}\otimes\mathds{1}_{A}],iH_{k}\rangle,  
\end{align}
 where we have made use of the fact that the cost function above can be obtained from \eqref{Eq:J} by setting $H_{p}=\mathds{1}-\ket{\psi_{T}}\bra{\psi_{T}}\otimes \mathds{1}_{A}$ where $\mathds{1}_{A}$ denotes the identity operator on subsystem $A$. Decreasing $J_{k}$ in each step can now be achieved by moving into the negative direction of the gradient given by Eq. \eqref{eq:opensysgradient}. 

If the auxiliary qubits are initially in a pure state, then due to the Stinespring dilation \cite{nielsen2002quantum}, there exists a unitary transformation over the composite system that allows for creating every state for the system qubits in subsystem $S$ as long as subsystem $A$ has dimension at most $d_{S}^{2}$, where $d_{S}$ is the dimension of subsystem $S$. For uniformly randomized $H_{k}$, the gradient can only vanish (with probability 1) when $\text{grad}J[U_{k}]=[\rho_{k}^{SA},\ket{\psi_{T}}\bra{\psi_{T}}\otimes\mathds{1}_{A}]=0$, i.e., at critical points that are given by saddle points and global optima, as in the closed system case \cite{Wu_2008}. Thus, when the initial system state $\rho_{0}^{S}$ does not commute with the target system state $\ket{\psi_{T}}\bra{\psi_{T}}$, we expect to obtain convergence almost surely to the target state. Although the fully mixed state $\rho_{0}^{S}=\mathds{1}/d_{A}$ trivially commutes with every target state, if we additionally apply a random unitary transformation to $\rho_0^{SA}$, we can almost surely obtain convergence to a generic pure state. Thus, randomized adaptive quantum state preparation allows for ``cooling'' the system from a state of infinite temperature, i.e., the fully mixed state, to a pure state of zero temperature by adaptively ``dumping'' entropy into the auxiliary system.

\section{Conclusion} 

We have introduced an algorithm for preparing quantum states that has favorable convergence properties and is applicable to almost all initial states. Knowledge of the initial and target quantum states is not required. The algorithm leverages randomization as the primary innovation, and operates by minimizing a cost function through an adaptively constructed quantum circuit. Each adaptive update step is informed by gradient measurements in which the associated tangent space directions are randomized. We have presented lower bounds on the average improvement that is obtained in each step, and have numerically studied the behavior of the algorithm for different randomization methods. We additionally discussed a generalization to mixed states that could be leveraged for thermal state preparation, an application area which is currently receiving significant interest \cite{PhysRevLett.123.220502, https://doi.org/10.48550/arxiv.1903.01451,NatBrandao,PhysRevApplied.16.054035,https://doi.org/10.48550/arxiv.2203.12757,https://doi.org/10.48550/arxiv.2002.00055,https://doi.org/10.48550/arxiv.2210.16419}.

A tradeoff is that, on one hand, the consideration of a uniformly random tangent space direction in each step allows for convergence to the target state, while on the other hand, the gradients may become exponentially small in the system size, thereby causing the convergence time to diverge and making the algorithm impractical for large-scale problems. This should not be surprising, as selecting directions at random is far from being optimal. The largest gradient, and thereby the largest (guaranteed) cost function change, is obtained when $iH_{k}=\text{grad}J[U_{k}]$. This situation corresponds to the full gradient flow, which is, in general, not efficiently implementable. It is an interesting question how gradient flows can be efficiently approximated \cite{https://doi.org/10.48550/arxiv.2202.06976} while maintaining convergence. To that end, the algorithm we present could be modified to project into random subspaces, rather than into a single random direction.

Furthermore, we emphasize that the algorithm possesses significant design flexibility that can be harnessed to increase its practicality, for example, by developing application-specific adaptations that tailor the randomness to the problem instance, e.g., by taking into account the symmetries of the target state \cite{gard2020efficient,https://doi.org/10.48550/arxiv.2109.05340,doi:10.1063/1.5110682,Anselmetti_2021}, and by studying how much the rate of convergence can be improved by incorporating classical optimization in different ways \cite{grimsley2019adaptive}. More fundamentally, we view the randomized adaptive circuit update, which can be applied to arbitrary input states and is depicted in Fig. \ref{Fig:Intro}, as a \emph{subroutine} that could be incorporated into or appended onto other algorithms \cite{https://doi.org/10.48550/arxiv.quant-ph/0001106,PhysRevLett.102.130503} in a flexible way, e.g., those utilizing problem-specific ans\"atze that aren't random, to improve convergence and state preparation fidelities.

We conclude by noting that we have focused this work on randomized adaptive quantum state preparation in the circuit model of quantum computing. However, randomness and adaptive constructions can also be created outside the circuit model, e.g. through random fields \cite{PhysRevX.7.041015,PhysRevLett.124.010405}. Extensions to this latter setting could open up new approaches for state preparation in both closed and open quantum systems that can be driven by an applied field, including qubit systems, analog quantum simulators, molecules, and materials.

\begin{acknowledgments}
\emph{Acknowledgements}.--- We gratefully acknowledge feedback from Mohan Sarovar and Andrew Baczewski on this work. We also would like to thank Thomas Schulte-Herbr\"uggen, Gunther Dirr, and Emanuel Malvetti for fruitful discussions. A.B.M. acknowledges support from Sandia National Laboratories’ Laboratory Directed Research and Development Program under the Truman Fellowship. Sandia National Laboratories is a multimission laboratory managed and operated by National Technology \& Engineering Solutions of Sandia, LLC, a wholly owned subsidiary of Honeywell International Inc., for the U.S. Department of Energy’s National Nuclear Security Administration under contract DE-NA0003525. This article has been authored by an employee of National Technology \& Engineering Solutions of Sandia, LLC under Contract No. DE-NA0003525 with the U.S. Department of Energy (DOE). The employee owns all right, title and interest in and to the article and is solely responsible for its contents. The United States Government retains and the publisher, by accepting the article for publication, acknowledges that the United States Government retains a non-exclusive, paid-up, irrevocable, world-wide license to publish or reproduce the published form of this article or allow others to do so, for United States Government purposes. The DOE will provide public access to these results of federally sponsored research in accordance with the DOE Public Access Plan https://www.energy.gov/downloads/doe-public-access-plan. This paper describes objective technical results and analysis. Any subjective views or opinions that might be expressed in the paper do not necessarily represent the views of the U.S. Department of Energy or the United States Government. C. A. and S. E. acknowledge support from the National Science Foundation (Grant No. 2231328). 
\end{acknowledgments}

\bibliography{bib}

\onecolumngrid

\appendix

\section{Lower bounding the cost function change} \label{sec:appA}
We consider
\begin{align}
J(x)=\bra{\psi_{k}}e^{ix H_{k}}H_{p}e^{-ix H_{k}}\ket{\psi_{k}}, 
\end{align}
and note that $J(x=0)=\bra{\psi_{k}}H_{p}\ket{\psi_{k}}=J_{k}$ and $J\left(x=-\gamma\frac{dJ(0)}{d\theta}\right)=J_{k+1}$. 

\subsubsection{Recalling results from classical optimization}  
In order to lower bound the cost function change $\Delta J_{k}=J_{k}-J_{k+1}$, we first recall some standard tools from classical optimization, particularly the gradient descent algorithm. For a continuously differentiable cost function $J(x)$, $x\in\mathbb R$, with Lipschitz continuous gradient $\frac{\partial J(x)}{\partial x}$, a direct application of Taylor's theorem  \cite{beck2017first} yields,  
\begin{align}
\label{eq:eineq1}
J(y)\leq J(x)+(y-x)\frac{\partial J(x)}{\partial x}+\frac{L_{f}^{2}}{2}(y-x)^{2}, 
\end{align}
where $L_{f}$ is the Lipschitz constant satisfying, 
\begin{align}
\frac{\left|\frac{\partial J(x)}{\partial x}-\frac{\partial J(y)}{\partial y}\right|}{|x-y|}\leq L_{f}.
\end{align}
Setting $x=0$ and $y=-\gamma\frac{\partial J(x)}{\partial x}|_{x=0}$ in \eqref{eq:eineq1} and picking $\gamma=\frac{1}{L_{f}}$ gives   \begin{align}
J_{k}-J_{k+1}\geq \frac{1}{2L_{f}}\left(\frac{\partial J(0)}{\partial x} \right)^{2},
\end{align}
which is a well known result for gradient algorithms \cite{beck2017first}. Thus, in order to lower bound the cost function change $\Delta J_{k}$, it remains to determine the Lipschitz constant $L_{f}$ for our setting.

\subsubsection{Determination of the Lipschitz constant} 
We first note that  $\frac{\partial J(x)}{\partial x}=\bra{\psi_{k}}e^{ix H_{k}}[iH_{k},H_{p}]e^{-ix H_{k}}\ket{\psi_{k}}\equiv \bra{\psi(x)}A\ket{\psi(x)}$, where we defined $\ket{\psi(x)}=e^{-ix H_{k}}\ket{\psi_{k}}$  and $A=[iH_{k},H_{p}]$. We can than upper bound $\left |\frac{\partial J(x)}{\partial x}-\frac{\partial J(y)}{\partial y}\right|$ by 
\begin{align}
\left |\frac{\partial J(x)}{\partial x}-\frac{\partial J(y)}{\partial y}\right|&=| \bra{\psi(x)}A\ket{\psi(x)}-\bra{\psi(y)}A\ket{\psi(y)}| \\
&=\frac{1}{2}\left |\big(\bra{\psi(x)}+\bra{\psi(y)}\big)A\big(\ket{\psi(x)}-\ket{\psi(y)}\big)+\big(\bra{\psi(x)}-\bra{\psi(y)}\big)A\big(\ket{\psi(x)}+\ket{\psi(y)}\big)\right |	\\
&\leq  2\Vert A \big(\ket{\psi(x)}-\ket{\psi(y)}\big)\Vert_{2}\leq 2 \Vert A\Vert_{2}\Vert \ket{\psi(x)}-\ket{\psi(y)}\Vert_{2}\\
&= 2 \Vert A\Vert_{2}\Vert (U_{x}-U_{y})\ket{\psi_{k}}\Vert_{2}=2 \Vert A\Vert_{2}\Vert(U_{y}^{\dagger}U_{x}-\mathds{1})\ket{\psi_{k}} \Vert_{2},
\end{align}
where we used unitary invariance of the vector norm, $U_{x}(s)=e^{-i x s H_{k}}$ and $U_{y}(s)=e^{-i y s H_{k}}$, using the short-hand notation $U_{x}(1)=U_{x}$, $U_{y}(1)=U_{y}$, and $\Vert A \Vert_{2}$ and $\Vert \ket{\psi} \Vert_{2}$ denote the spectral norm of $A$ and the Euclidean vector norm of $\ket{\psi}$, respectively. Since 
\begin{align}
U_{y}^{\dagger}U_{x}-\mathds{1}=i\int_{0}^{1}ds\, U_{y}^{\dagger}(s)(y-x)H_{k}U_{x}(s), 	
\end{align}
 we have 
\begin{align}
\Vert(U_{y}^{\dagger}U_{x}-\mathds{1})\ket{\psi_{k}} \Vert_{2}&\leq \int_{0}^{1}ds\,\Vert (U_{y}^{\dagger}(s)(y-x)H_{k}U_{x}(s))\ket{\psi_{k}} \Vert_{2}	\\
&=|x-y|\int_{0}^{1}ds\,\Vert H_{k}\Vert_{2}\Vert \ket{\psi(x)}\Vert_{2}=|x-y|\Vert H_{k}\Vert_{2}=|x-y|\Vert H\Vert_{2}.
\end{align}
In the last step we used unitary invariance of the spectral norm, assuming that $H_{k}=V_{k}^{\dagger}HV_{k}$ where $V_{k}$ is unitary and $H$ is a Hermitian matrix. Since
\begin{align}
\Vert A\Vert_{2}=\Vert [H_{k}, H_{p}]\Vert_{2}\leq 2 \Vert H_{k}\Vert_{2}\Vert H_{p}\Vert_{2}=2\Vert H_{p}\Vert_{2}\Vert H \Vert_{2}, 	
\end{align}
we arrive at 
\begin{align}	
\frac{\left |\frac{\partial J}{\partial x}-\frac{\partial J}{\partial y}\right|}{|x-y|} \leq 4 \Vert H_{p} \Vert_{2}\Vert H \Vert_{2}^{2}, 
\end{align}
from which we conclude that the Lipschitz constant is given by $L_{f}=4 \Vert H_{p} \Vert_{2}\Vert H \Vert_{2}^{2}$.  Further assuming $\Vert H\Vert_{2}=1$, we arrive at the lower bound \eqref{Eq:ChangeinJ} given in the paper. 

\section{Expected cost function change when sampling from unitary 2-designs} \label{sec:appB}

We first note that the expected value of the gradient $\mathbb E_{V_{k}}\langle \text{grad}J[U_{k}],iH_{k}\rangle$ is given by  
\begin{align}
\mathbb E_{V_{k}}\langle \text{grad}J[U_{k}],iV_{k}^{\dagger}HV_{k}\rangle=\langle \text{grad}J[U_{k}],i\mathds{1}\rangle\frac{\text{Tr}\{H\}}{d}=0, 
\end{align}
when $V_k$ is sampled according to the Haar measure on $\text{SU}(d)$ or from a unitary 2-design. Consequently, in this case $\mathbb E_{V_{k}}\langle \text{grad}J[U_{k}],iH_{k}\rangle^{2}$ is equal to the variance.  

In order to lower bound the expected cost function change $\mathbb E_{V_{k}}\Delta J_{k}$ when $V_{k}$ is drawn according to the Haar measure on $\text{SU}(d)$ or sampled from a unitary 2-design, we first note that from the lower bound \eqref{Eq:ChangeinJ} we have $\mathbb E_{V_{k}}\Delta J_{k}\geq \frac{1}{8\Vert H_{p}\Vert_{2}}\mathbb E_{V_{k}}\langle \text{grad}J[U_{k}],iH_{k}\rangle^{2}=\frac{1}{8\Vert H_{p}\Vert_{2}}\mathbb E_{V_{k}}\bra{\psi_{k}}[iV_{k}^{\dagger}HV_{k},H_{p}]\ket{\psi_{k}}^{2}$.To calculate the expectation on the right-hand side, we make use of the relation \cite{BarrenPoly},
\begin{align}
\int_{V\in\text{SU}(d)}\,dV\,\text{Tr}\{V^{\dagger}BVCV^{\dagger}DVA\}&=\frac{1}{d^{2}-1}\left(\text{Tr}\{A\}\text{Tr}\{C\}\text{Tr}\{BD\}+\text{Tr}\{AC\}\text{Tr}\{B\}\text{Tr}\{D\}\right)\nonumber \\
&-\frac{1}{d(d^{2}-1)}\left(\text{Tr}\{AC\}\text{Tr}\{BD\}+\text{Tr}\{A\}\text{Tr}\{B\}\text{Tr}\{C\}\text{Tr}\{D\}\right).
\end{align}
We find, 
\begin{align}
\mathbb E_{V_{k}}\bra{\psi_{k}}[iV_{k}^{\dagger}HV_{k},H_{p}]\ket{\psi_{k}}^{2}
&=\mathbb E_{V_{k}}\text{Tr}\{[iV_{k}^{\dagger}HV_{k},H_{p}]\ket{\psi_{k}}\bra{\psi_{k}}[iV_{k}^{\dagger}HV_{k},H_{p}]\ket{\psi_{k}}\bra{\psi_{k}}\}, \nonumber \\
&=-2\text{Tr}\{H^{2}\}\left(\frac{1}{d^{2}-1}\bra{\psi_{k}}H_{p}\ket{\psi_{k}}^{2}-\frac{1}{d(d^{2}-1)}\bra{\psi_{k}}H_{p}\ket{\psi_{k}}^{2}\right.\nonumber\\
&\left.-\frac{1}{d^{2}-1}\bra{\psi_{k}}H_{p}^{2}\ket{\psi_{k}}+\frac{1}{d(d^{2}-1)}\bra{\psi_{k}}H_{p}\ket{\psi_{k}}^{2}\right)\nonumber \\
&=\frac{2\text{Tr}\{H^{2}\}}{d^{2}-1}\left(\bra{\psi_{k}}H_{p}^{2}\ket{\psi_{k}}-\bra{\psi_{k}}H_{p}\ket{\psi_{k}}^{2}\right),
\end{align}
which yields the desired result given in \eqref{eq:lowerboundHaar} in the paper.

\end{document}